\documentclass[twocolumn,twoside,slac_two]{revtex4}
\usepackage{graphicx}
\usepackage{fancyhdr}
\begin{document}

\title{Insights into the particle acceleration of a peculiar gamma-ray radio galaxy IC 310}

%

\author{J. Sitarek}
\affiliation{University of \L\'od\'z, PL-90236 Lodz, Poland, \\
IFAE, Campus UAB, E-08193 Bellaterra, Spain}
\author{D. Eisenacher Glawion, K. Mannheim}
\affiliation{Universit\"at W\"urzburg, D-97074 W\"urzburg, Germany}
\author{P. Colin for the MAGIC Collaboration}
\affiliation{Max-Planck-Institut f\"ur Physik, M\"unchen, Germany}
\author{M. Kadler}
\affiliation{Universit\"at W\"urzburg, D-97074 W\"urzburg, Germany }
\author{R. Schultz, F. Krau\ss}
\affiliation{Universit\"at W\"urzburg, D-97074 W\"urzburg, Germany,\\
Dr. Remeis-Sternwarte Bamberg, Astronomisches Institut der Universit\"at Erlangen-N\"urnberg, ECAP, D-96049 Bamberg, Germany}
\author{E. Ros}
\affiliation{Max-Planck-Institut f\"ur Radioastronomie, D-53121 Bonn, German,\\
Observatori Astron\`omic, Universitat de Val\`encia, E-46980 Paterna, Val\`encia, Spain,\\
Departament d'Astronomia i Astrof\'{\i}sica, Universitat de Val\`encia, E-46100 Burjassot, Val\`encia, Spain}
\author{U. Bach}
\affiliation{Max-Planck-Institut f\"ur Radioastronomie, D-53121 Bonn, German}
\author{J. Wilms}
\affiliation{Dr. Remeis-Sternwarte Bamberg, Astronomisches Institut der Universit\"at Erlangen-N\"urnberg, ECAP, D-96049 Bamberg, Germany}

\begin{abstract}
IC 310 has recently been identified as a gamma-ray emitter based on observations at GeV energies with Fermi-LAT and at very high energies (VHE, $E > 100$ GeV) with the MAGIC telescopes. 
Despite IC 310 having been classified as a radio galaxy with the jet observed at an angle $> 10$ degrees, it exhibits a mixture of multiwavelength properties of a radio galaxy and a blazar, possibly making it a transitional object.
On the night of 12/13$^\mathrm{th}$ of November 2012 the MAGIC telescopes observed a series of violent outbursts from the direction of IC 310 with flux-doubling time scales faster than 5 min and a peculiar spectrum spreading over 2 orders of magnitude. 
Such fast variability constrains the size of the emission region to be smaller than 20\% of the gravitational radius of its central black hole, challenging the shock acceleration models, commonly used in explanation of gamma-ray radiation from active galaxies. 
Here we will show that this emission can be associated with pulsar-like particle acceleration by the electric field across a magnetospheric gap at the base of the jet.
\end{abstract}

\maketitle

\thispagestyle{fancy}


\section{INTRODUCTION}

The nearby lenticular (S0, $z=0.0189$) galaxy IC\,310 located in the Perseus cluster exhibits an active galactic 
nucleus (AGN). 
This object has been detected at high energies (above 30\,GeV) with \textit{Fermi}/LAT \cite{nsv10} as well as at TeV energies \cite{al10,al13}.
The jet of IC310, extending in the outward direction from the center of the cluster led to early assignment of this object as a head-tail radio galaxy \cite{ryle68, sijbring, miley80}. 
However, using the Very-Long-Baseline Interferometry (VLBI) technique, a parsec-scale one-sided jet was found to follow  the large-scale jet within about $10^\circ$ \cite{kadler12}.
The alignment of the jet at different scales, without any hints of bending put in doubt the above classification.
Instead, the inner jet appears to be blazar-like with a missing counter jet due to relativistically boosted emission.
Further indications for transitional behavior between a radio galaxy and a blazar were found in IC\,310 in various energy ranges \cite{rector}.
The mass, $M$,  of the black hole of IC\,310 can be inferred from its relation with the velocity dispersion, $\sigma$, of the host galaxy \cite{gultekin2009, al14}, namely $M=(3^{+4}_{-2}) \times10^{8}M_\odot$.  

MAGIC (Major Atmospheric Gamma Imaging Cherenkov) is a system of two 17-m diameter Imaging Atmospheric Cherenkov telescopes located on La Palma, Canary Islands. 
It allows observations of gamma-ray sources with energies above 50\,GeV.
During the observations of the Perseus cluster performed in the end of 2012 MAGIC telescopes revealed an extreme gamma-ray flare from IC\,310 on the night of 12/13$^\mathrm{th}$ of November \cite{al14}.
In addition, the source was observed in radio band by European VLBI Network (EVN) during October/November 2012. 

In Section~\ref{sec:results} we report the data analysis and results of the MAGIC observations during the flare and the radio observations. 
In Section~\ref{sec:interp} we discuss possible interpretation of the ultrafast variability of the gamma-ray emission observed from IC\,310. 

\section{RESULTS}\label{sec:results}
\subsection{MAGIC}
MAGIC telescopes were observing the Perseus cluster on the night of 12/13$^\mathrm{th}$ of November for 3.7\,h. 
The observations consisted of 4 pointings, two of the them with a standard offset of 0.4$^{\circ}$ with respect to IC\,310 and the remaining ones are at a distance $0.94^{\circ}$ away from the object.
The signal extraction and calibration of the data, the image parametrization, the direction and energy reconstruction as well as the gamma-hadron separation were applied with the standard analysis software MARS as described in \cite{zanin13}. 

In the night of the flare a strong signal of 507 gamma-like events above 300\,GeV in the region around IC\,310 in excess of the background estimated as 47 events was observed. 
Due to still limited statistics of events and a very rapid variability behavior, the classical approach for the calculation of light curves in gamma-ray astronomy which is based on the fixed width of the time bins is not optimal in this case. 
We used instead a method similar to the one commonly used for data of X-ray observatories for the computing of energy spectra. 
We first identify all periods in the data during which the telescopes were not operational (in particular $\lesssim 1$\,min gaps every 20\,min when the telescope is slewing and reconfiguring for the next data run). 
Afterwards, we bin the remaining time periods based on a fixed number (in this case 9) of ON events per bin. 
We estimate the number of background events in each time bin from four off-source regions at the same distance from the camera center.
Using toy MC simulations we validated that this method limits the bias in flux value and its error \cite{al14}. 
As the signal to background ratio above 300\,GeV is much larger than 1 this assures that the precision of individual points in the light curve is close to $3\sigma$. 

\begin{figure}[t]
\centering
\includegraphics[width=0.49\textwidth]{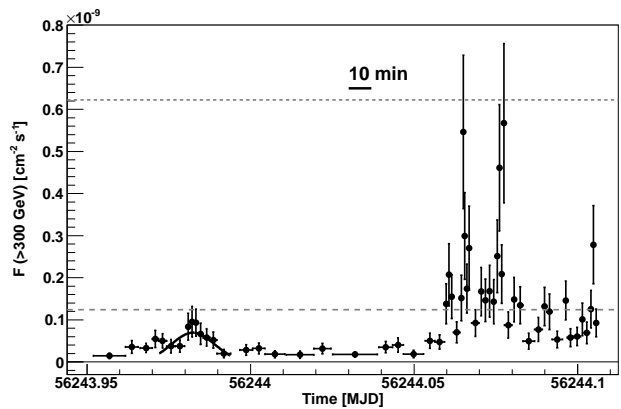}
\caption{
Light curve of IC~310 observed with the MAGIC telescopes in the night of November 12/13$^{\mathrm{th}}$, 2012, above 300\,GeV. 
As a flux reference, the two gray lines indicate levels of $1$ and $5$ times the flux level of the Crab Nebula, respectively.
The precursor flare (MJD 56243.972--56243.994) has been fitted with a Gaussian distribution.
The figure is reprinted from \cite{al14}.
}
\label{Lightcurve}%
\end{figure}

The resulting light curve is presented in Fig.~\ref{Lightcurve}.
The mean flux above 300\,GeV during this period is $\Phi_{\mathrm{mean}}=(6.1 \pm 0.3)\times10^{-11}$\,cm$^{-2}$s$^{-1}$. 
This is four times higher than the high state flux of $(1.60 \pm 0.17)\times10^{-11}$\,cm$^{-2}$s$^{-1}$ reported in \cite{al13}.
The emission is highly variable, fitting the light curve in the full time range with a constant reveals a $\chi^2/\mathrm{N.d.o.f}$ of 199/58 corresponding to a probability of $2.6\times10^{-17}$. 

\begin{figure}
   \centering
      \includegraphics[width=0.49\textwidth]{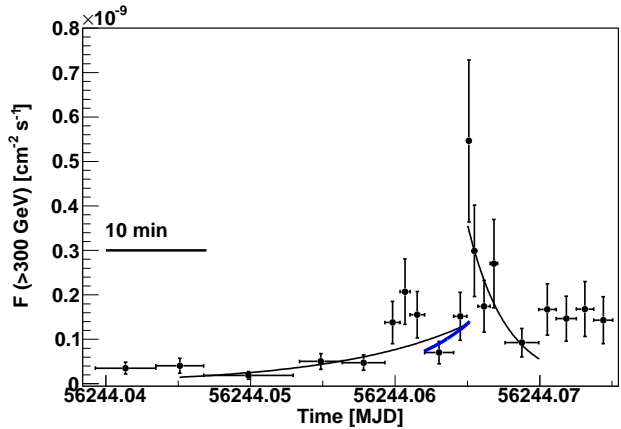}
   \caption{Zoom of the first big flare seen in the light curve of IC\,310  above 300\,GeV. 
Black lines show exponential fits to the rising and decay edges to the substructures in the light curve. 
The blue line shows the slowest doubling time necessary to explain the raising part of the flare at C.L. of 95\%. 
The figure is reprinted from \cite{al14}.
}
\label{figS4}
\end{figure}

We use the rapidly rising part of the 1$^{\mathrm{st}}$ big flare (MJD 56244.0620--56244.0652) in order to compute the
conservative, slowest doubling time, $\tau_\mathrm{D}$, which is still consistent with the MAGIC data.
We fit the light curve with a set of exponential functions, each time assuming a given $\tau_\mathrm{D}$ value and computing the corresponding fit probability.
We obtain that $4.9\,\mathrm{min}$ is the largest value of $\tau_\mathrm{D}$, which can still marginally fit the data with probability $>5\%$ (see the blue line in Fig.~\ref{figS4}).
Note that the corresponding time scale in the frame of reference of IC 310 will be slightly shorter: $4.9/(1+z)\,\mathrm{min} = 4.8\,\mathrm{min}$.

\begin{figure}
  \centering
  \includegraphics[width=0.49\textwidth]{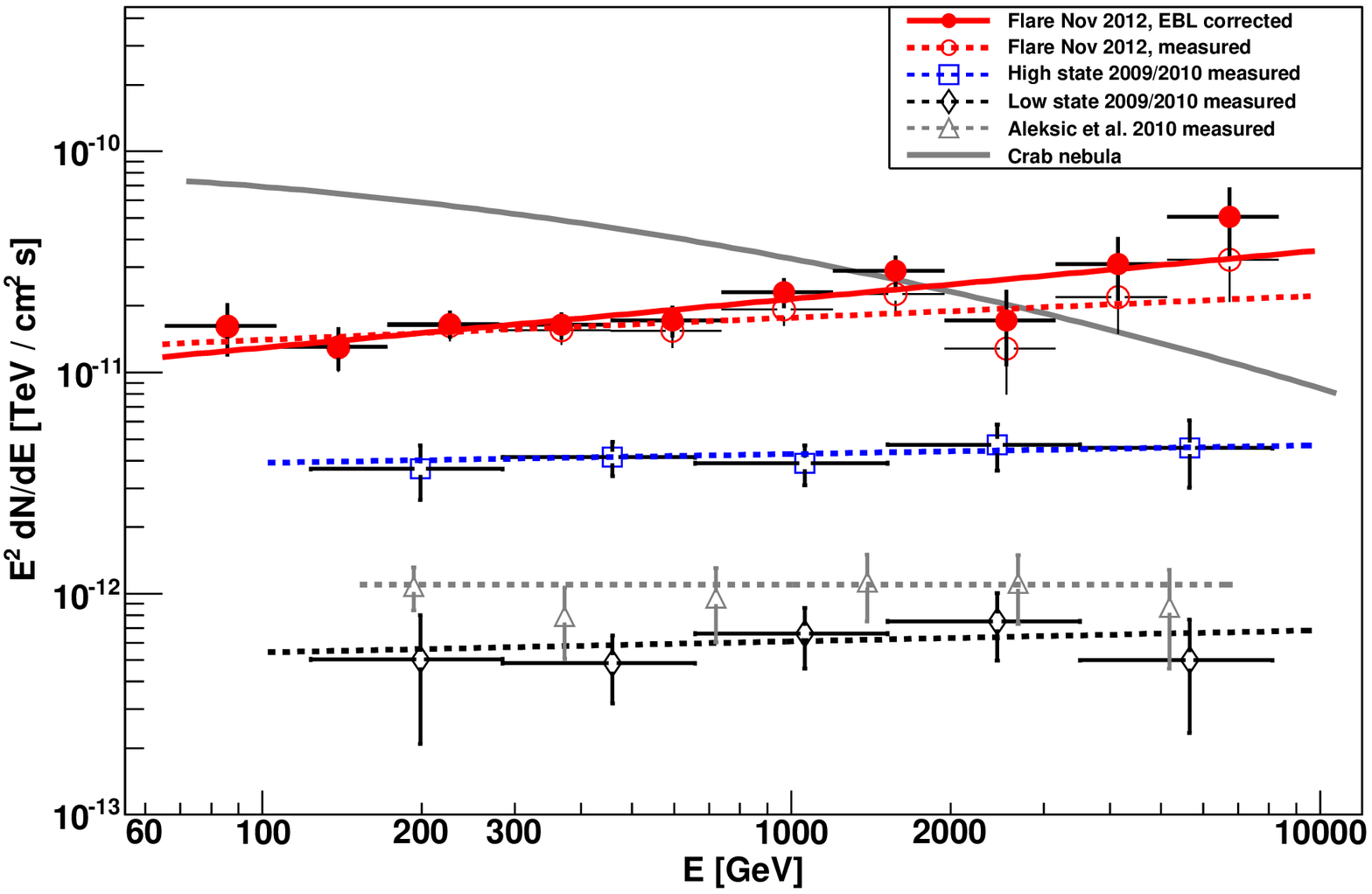} 
  \caption{
MAGIC measurement of average spectral energy distributions of IC~310 during the flare (red) .  
For comparison we show the results from the high (blue, open squares) and low (black, open markers) states reported in \cite{al13}  and the average results (gray triangles) reported in \cite{al10}. 
The dashed lines show power-law fits to the measured spectra, and the solid line with filled circles depicts the spectrum corrected for absorption in the extragalactic background light according to \cite{dominguez11}. 
 As a reference, the spectral power-law fit of the Crab Nebula observations from \cite{al12} is shown (gray, solid line). 
The figure is reprinted from \cite{al14}.
}
  \label{SED}%
\end{figure}

The observed spectrum can be described by a simple power law (see Fig.~\ref{SED}):
\begin{equation}
\frac{\mathrm{d}F}{\mathrm{d}E}=f_0\times\left(\frac{E}{1\mathrm{TeV}}\right)^{-\Gamma}.
\end{equation}
The flux normalization at 1\,TeV obtained from the fit is $f_0=(17.7\pm0.9_{\rm stat}\pm2.1_{\rm syst}) \times10^{-12} \mathrm{TeV}^{-1}\,\mathrm{cm}^{-2}\,\mathrm{s}^{-1}$ 
Even while the mean flux during the flaring night is $4-30$ larger than previous measurements, the spectral index, $\Gamma = 1.90\pm0.04_{\rm stat}\pm0.15_{\rm syst}$ is consistent with them within the statistical and systematic errors.
No significant bend or cut-off is seen in the spectrum up to TeV energies. 
As part of the observation was carried out with a higher then usual offset angle from the camera center the systematic error on the flux normalization is slightly larger (12\%) then reported in \cite{al14b}. 
The error of the energy scale is 15\%. 

\subsection{EVN}

IC\,310 has been observed with the EVN at 1.7, 5.0, 8.4 and 22.2\,GHz between 2012-10-21 and 2012-11-07.
The data were amplitude and phase calibrated using standard procedures with the Astronomical Image Processing System (\textsc{AIPS}, \cite{Greisen2003}) and imaged and self-calibrated using \textsc{DIFMAP} \cite{Shepherd1994}. 

\begin{figure}
\centering
\includegraphics[width=0.49\textwidth]{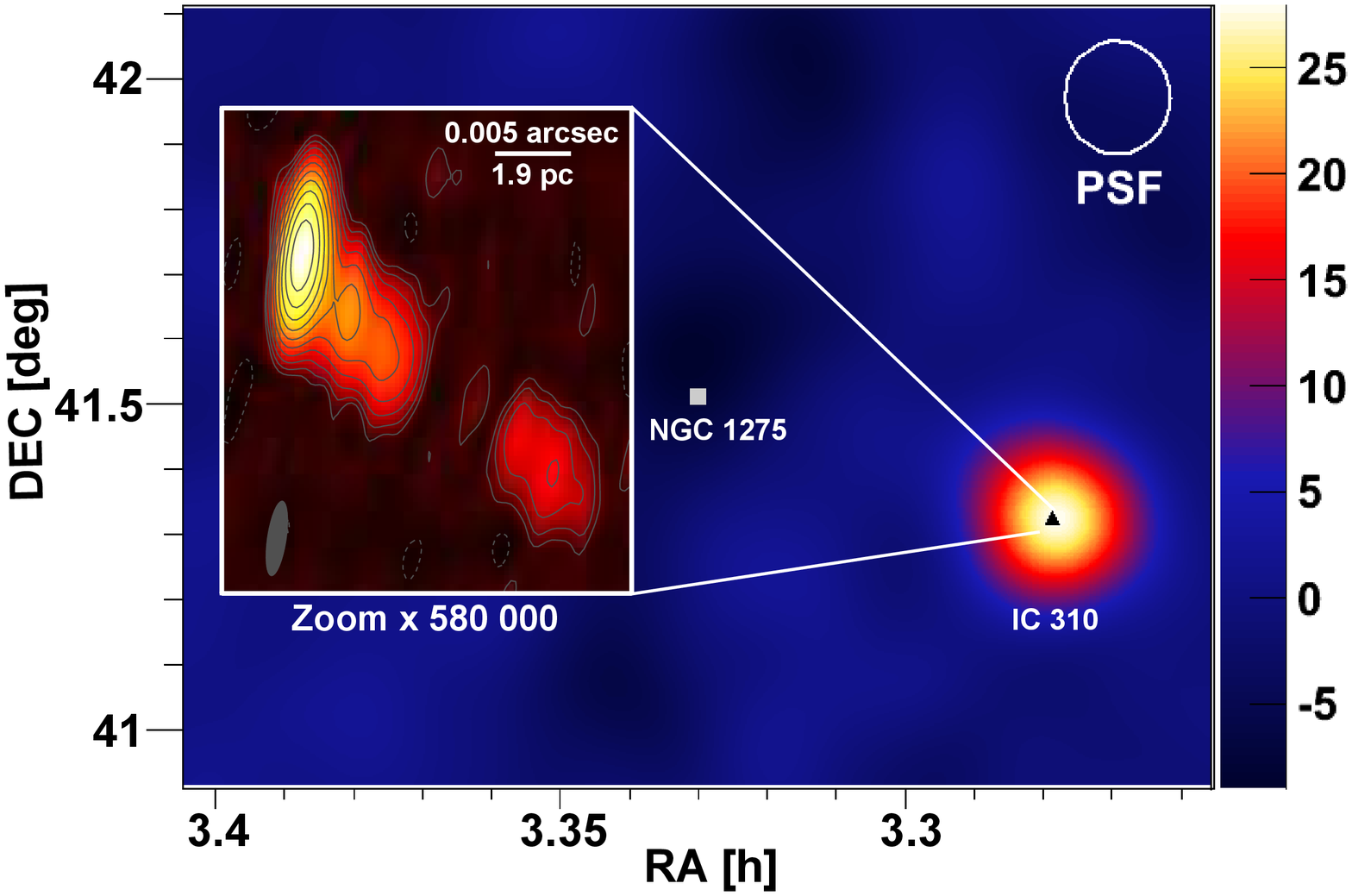}
\caption{
Significance map (color scale) of the Perseus cluster in gamma rays observed in the night of November 12/13$^{\mathrm{th}}$, 2012, with the MAGIC telescopes. 
The inset shows the radio jet image of IC\,310 at 5.0\,GHz obtained with the European VLBI Network (EVN) on October 29, 2012. 
Contour lines (and associated to them color scale) increase logarithmically by factors of 2 starting at three times the noise level.
The ratio of the angular resolution between MAGIC and the EVN is 1:580\,000.
The figure is reprinted from \cite{al14}.
}
\label{Skymap}%
\end{figure}

In inset panel of Fig.~\ref{Skymap} we present the image with the highest dynamic range obtained from the observation at 5.0 GHz from 2012-10-29. 
The image has a peak flux density of 77\,mJy/beam and a 1$\sigma$ noise level of 0.027\,mJy/beam.
The restoring beam has a major and minor axis of $4.97\times1.24$\,mas$^{2}$ with the major axis at a position angle of $-8.5^{\circ}$. 
It contains a total flux density of $S_\mathrm{total}=109\mathrm{\,mJy}$, which we conservatively assume to be 
accurate to 10\%. 
The dynamic range $DR$ of the image, defined as the ratio of the peak flux density to tripled noise level in the image is $\approx 950$. 

The angle $\theta$ of the radio jet to the line-of-sight can be determined from Doppler boosting arguments for a given jet speed $\beta$ and spectral 
index $\alpha$ by considering the ratio $R$ of the flux density in the jet and counter-jet:
\begin{equation}
 R=\left(\frac{1+\beta\cos\theta}{1-\beta\cos\theta}\right)^{2-\alpha}. \label{eq1}
\end{equation}
Following \cite{kadler12} we use the $DR$ as an upper limit for the detection of a counter-jet. This gives us an upper limit of $\theta$:
\begin{equation}
 \theta < \mathrm{arc\,cos} \left(\frac{DR^{1/(2-\alpha)}-1}{DR^{1/(2-\alpha)}+1}\right).
\end{equation}
 
Substituting $DR$ in Eq.~\ref{eq1}, assuming a flat spectral index of $\alpha=0$ and
$\beta\rightarrow1$, we obtain an upper limit for the angle between the jet and the line-of-sight of $\lesssim 20^{\circ}$.

Additionally, the extension of the projected one-sided kpc radio jet of $\sim350$\,kpc measured at a wavelength of 49\,cm \cite{sijbring} yields an estimate of a lower limit for the angle.
De-projecting the jet using the upper limit quoted above would results in a lower limit of the jet length of $\sim$1\,Mpc. 
Radio galaxies typically show jets extending up to 150\,kpc-300\,kpc \cite{neeser95}. 
The maximal length of radio jets has been measured to be a few Mpcs which corresponds to an angle of $\sim5-10^{\circ}$ in the case of IC\,310. 
Smaller angles would rapidly increase the de-projected length of the jet to values far above the maximum of the distribution of the jet lengths. 

\section{INTERPRETATION}\label{sec:interp}
GeV and TeV gamma-ray emission from blazars and radio galaxies is often explained in terms of shock-in-jet models. 
Charged particles are accelerated in an active region moving along the jet.
Causality condition provides that the variability time scale of the observed emission can be used to constrain the size of the emission region.

A conservative estimate of the shortest variability time scale in the frame of reference of IC\,310 yields $\Delta t/(1+z)=4.8$~min.
Using the best mass estimate of IC\,310 black hole this measurement corresponds to $20\%$ of the light travel time across the event horizon.
Even allowing for the factor 3 uncertainty in the mass the fraction, $60\%$, is still below one.
The ultrafast variability casts a shadow of doubt on the current shock-in-jet paradigm.
The moving shock plasma leads to a shortening of the observed variability time scale $\Delta t$ compared with the
variability time scale $\Delta t'$ in a frame comoving with the shock given by 
$\Delta t=(1+z)\delta^{-1}\Delta t'$.
This effect is often used to explain ultrafast variability from blazars \cite{albert07a, aharonian07} in which $\delta$ can be nearly arbitrarily large providing that the jet moving with large Lorentz factor is observed at a very small angle. 
In the case of IC\,310 however the estimation of the observation angle $10^\circ-20^\circ$ obtained from the radio observations constrain the maximum Doppler factor to be $\lesssim 6$.

All of these attempts to explain the sub-horizon scale variability with relativistic projection effects alone encounter a fundamental problem \cite{np12}. 
If the perturbations giving rise to the blazar variability are injected at the jet base, the time scale of the flux variations in the frame comoving with the jet is affected by time dilation with Lorentz factor $\Gamma_{\rm j}$.  
In blazars where $\delta\sim\Gamma_{\rm j}$, the Lorentz factor cancels out, and the observed variability time scale is ultimately bounded below by $\Delta t_{\rm BH}$. 

Additionally, a very high value of the  Doppler factor is required to avoid the absorption of the TeV gamma rays due to interactions with low-energy synchrotron photons.
Such synchrotron photons are inevitably produced together with the gamma rays in the shock-in-jet scenario. The optical depth to pair creation by the gamma rays can be approximated by
$\tau_{\gamma\gamma}(10~\rm TeV) \sim 300 \left(\delta / 4\right)^{-6}\left (\Delta t / 1~min\right)^{-1} \left(L_{\rm syn} / 10^{42}~\rm erg~s^{-1}\right)$.  
Adopting, conservatively,  a non-thermal infrared luminosity of $\sim 1\%$ of the gamma-ray luminosity during the flare, the emission region would be transparent to the emission of 10~TeV gamma rays only if $\delta\gtrsim10$.

In summary, trying to interpret the IC\,310 flare in the framework of the shock-in-jet model meets difficulties.
Alternative models can involve stars falling into the jet \cite{bednarek97, barkov10}, mini-jet structures within the jets \cite{giannios10} or magnetospheric models \cite{rieger00, neronov07, levinson11, beskin92}.
In the case of IC\,310 star-in-jet model cannot provide sufficient luminosity to explain the TeV flare \cite{al14}.
Also jets-in-jet models suffer from rapidly dropping luminosity at larger observation angles \cite{al14}. 
Moreover the magnetic reconnection which can led to production of such mini-jets is expected to occur in the main jet rather at larger distances from the black hole.

\begin{figure}
\centering
\includegraphics[width=0.49\textwidth]{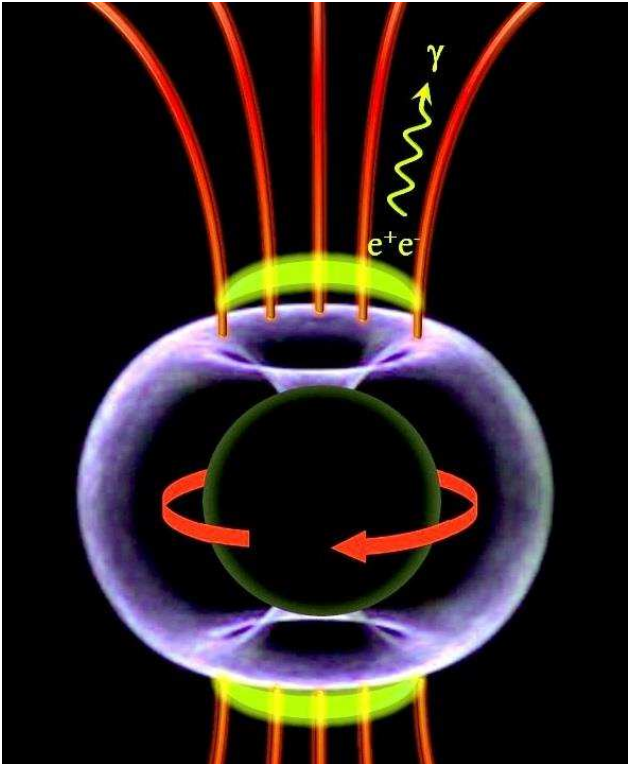}
\caption{
Scenario for the magnetospheric origin of the gamma-rays: 
A maximally rotating black hole with event horizon $r_{\rm g}$ (black sphere)
accretes plasma from the center of the galaxy IC~310.  In the apple-shaped ergosphere (blue) extending to $2 r_{\rm g}$ in the equatorial plane, 
Poynting flux is generated by the frame-dragging effect.
The rotation of the black hole induces a charge-separated magnetosphere (red)
with polar vacuum gap regions (yellow). In the gaps, the electric field of the
magnetosphere has a component  parallel to the magnetic field accelerating
particles to ultra-relativistic energies.  
Inverse-Compton scattering and intense pair production due to interactions with low-energy thermal photons from the plasma accreted by the black hole leads to the observed gamma rays.
The figure is reprinted from \cite{al14}.
}
\label{gap}%
\end{figure}

In magnetospheric models, particle acceleration is assumed occur in electric fields parallel to the magnetic fields. 
This mechanism is common to the particle-starved magnetospheres of pulsars, but it could also operate in the magnetospheres anchored to the ergospheres of accreting black holes (see Fig.~\ref{gap}).
Electric fields can exist in vacuum gaps when the density of charge carriers is too low to cause their shortcut, i.e. below the so-called Goldreich-Julian charge density.  
Electron-positron pairs in excess of the Goldreich-Julian charge density can be produced thermally by photon-photon collisions in a hot accretion torus or corona surrounding the black hole.
It has also been suggested, that particles can be injected by the reconnection of twisted magnetic loops in the accretion flow \cite{neronov09}.
A depletion of charges from thermal pair production is expected to happen when the accretion rate becomes very low.  
In this late phase of their accretion history, supermassive black holes are expected to have spun up to maximal rotation.  
Black holes can sustain a Poynting flux jet by virtue of the Blandford-Znajek mechanism \cite{blandfordznj77}. 
Jet collimation takes place rather far away from the black hole, i.e. at the scale of the light cylinder beyond $\sim 10r_{\rm g}$.
Gaps could be located at various angles with the jet axis corresponding to the polar and outer gaps in pulsar magnetospheres leading to fan beams at rather large angles with the jet axis.
As the gap height and seed particle content depend sensitively on plasma turbulence and accretion rate, the gap emission is expected to be highly variable.
For an accretion rate of $\sim 10^{-4}$ of the Eddington accretion rate and maximal black hole rotation, the gap height in IC~310 is expected to be $h\sim 0.2 r_{\rm g}$ \cite{levinson11} which is in line with the variability times seen in the observations.
Depending on the electron temperature and geometry of the radiatively inefficient accretion flow, its thermal cyclotron luminosity can be low enough to warrant the absence of pair creation attenuation in the spectrum of gamma rays. 
In this picture, the intermittent variability witnessed in IC~310 is due to a runaway effect.  
As particles accelerate to ultrahigh energies, electromagnetic cascades develop multiplying the number of charge carriers until their current shortcuts the gap.  
The excess particles are then swept away with the jet flow, until the gap reappears.

\section{CONCLUSIONS}

Radio galaxies and blazars with very low accretion rates allow us to obtain a glimpse of the jet formation process near supermassive black holes. 
Observations of IC\,310 performed with the MAGIC telescopes showed variability with time scale below 5\,min, shorter than the light crossing time of the event horizon of its black hole.
The commonly used in AGNs shock-in-jet models have troubles to explain such emission.
A plausible explanation involves emission from vacuum gaps in the magnetosphere of IC\,310. 
Interestingly, such explanation invite to explore analogies with pulsars where particle acceleration takes place in two stages. 
In the first stage, particle acceleration occurs in the gaps of a charge-separated magnetosphere anchored in the ergosphere of a rotating black hole, and in a second stage at shock waves in the force-free wind beyond the outer light cylinder.

\bigskip 
\begin{acknowledgments}
We would like to thank
the Instituto de Astrof\'{\i}sica de Canarias
for the excellent working conditions
at the Observatorio del Roque de los Muchachos in La Palma.
The support of the German BMBF and MPG,
the Italian INFN,
the Swiss National Fund SNF,
and the Spanish MICINN
is gratefully acknowledged.
This work was also supported
by the CPAN CSD2007-00042 and MultiDark CSD2009-00064 projects of the Spanish Consolider-Ingenio 2010 programme,
by grant 127740 of the Academy of Finland,
by the DFG Cluster of Excellence ``Origin and Structure of the Universe'',
by the Croatian Science Foundation (HrZZ) Projects 09/176,
by the University of Rijeka Project 13.12.1.3.02,
by the DFG Collaborative Research Centers SFB823/C4 and SFB876/C3,
and by the Polish MNiSzW grant 745/N-HESS-MAGIC/2010/0.
We thank also the support by DFG WI 1860/10-1.
J. S. was supported by ERDF and the Spanish MINECO through
FPA2012-39502 and JCI-2011-10019 grants.
E. R. was partially supported by the Spanish MINECO projects
AYA2009-13036-C02-02 and AYA2012-38491-C02-01 and by
the Generalitat Valenciana project PROMETEO/2009/104, as
well as by the COST MP0905 action 'Black Holes in a Violent
Universe'.
The European VLBI Network is a joint facility of European, Chinese,
South African and other radio astronomy institutes funded by their national research councils.
The research leading to these results has received funding from
the European Commission Seventh Framework Programme (FP/2007-2013)
under grant agreement No. 283393 (RadioNet3).
\end{acknowledgments}

\bigskip 

\end{document}